\documentclass[a4paper, amsfonts, amssymb, amsmath, reprint, showkeys, nofootinbib, twoside]{revtex4-1}
\usepackage[english]{babel}
\usepackage[utf8]{inputenc}
\usepackage[colorinlistoftodos, color=green!40, prependcaption]{todonotes}
\usepackage{amsthm}
\usepackage{mathtools}
\usepackage{physics}
\usepackage{xcolor}
\usepackage{graphicx}
\usepackage[left=23mm,right=13mm,top=35mm,columnsep=15pt]{geometry} 
\usepackage{adjustbox}
\usepackage{placeins}
\usepackage[T1]{fontenc}
\usepackage{lipsum}
\usepackage{csquotes}
\usepackage[pdftex, pdftitle={Article}, pdfauthor={Author}]{hyperref} 
\begin{document}
\title{Microwave Superconductivity}

\author{Steven M. Anlage}
    \email[Correspondence email address: ]{anlage@umd.edu}
    \affiliation{Quantum Materials Center, Physics Department, University of Maryland, College Park, MD 20742-4111  USA}

\date{\today} 

\begin{abstract}
We give a broad overview of the history of microwave superconductivity and explore the technological
developments that have followed from the unique electrodynamic properties of superconductors. Their low
loss properties enable resonators with high quality factors that can nevertheless handle extremely high
current densities. This in turn enables superconducting particle accelerators, high-performance filters and
analog electronics, including metamaterials, with extreme performance. The macroscopic quantum properties has enabled new generations of ultra-high-speed digital computing and extraordinarily sensitive detectors. The microscopic quantum properties have enabled large-scale
quantum computers, which at their heart are essentially microwave-fueled quantum engines. We celebrate the rich history
of microwave superconductivity and look to the promising future of this exciting branch of microwave technology.
\end{abstract}


\maketitle

\section{INTRODUCTION}
The unique microwave properties of superconductors enable a remarkable range of novel applications and technologies.  The low Ohmic losses of the superconducting state allow for extremely high efficiency and compact charged particle accelerators based on microwave resonant cavities with quality factors exceeding $10^{11}$.  The unique electrodynamic properties of superconductors enable low-dispersion transmission lines that preserve the integrity of extremely short electrical impulses.  This, along with the macroscopic quantum properties of superconductors have enabled a family of radically new digital electronics based on magnetic flux quantization and the Josephson effect.  The low-loss and nonlinear properties of superconductors create an ideal setting for demonstration of quantum phenomena such as entanglement and controlled quantum state evolution.  Because microwave superconductivity is a key enabler for present and future quantum technologies, anyone trained in microwave engineering has entry level skills for this exciting new technology frontier \cite{Zag11}.  My goal in this article is to give an overview of the remarkable microwave technologies uniquely enabled by superconductivity.  I will also argue that the barrier to utilizing these ‘exotic’ technologies has been considerably lowered in recent years through the development of inexpensive and highly reliable cryogenic technology infrastructure. \\

	This overview touches lightly on many fascinating topics.  The technical detail (and rigor) is kept to a minimum in order to bring out the main trends in the development of superconductors for microwave applications.  To delve deeper, the interested reader is advised to consult a number of helpful books, monographs, and articles on microwave superconductivity that have appeared in the recent past.  The most comprehensive is probably the collected papers from a NATO Advanced Study Institute \cite{Wein12}.  Other accessible but more specialized sources on microwave superconductivity include works on RF superconductivity for particle accelerator applications,\cite{Pad98,Pad17} Josephson junction dynamics,\cite{Lik86} electrodynamics of high-temperature superconductors,\cite{Port93} analog superconducting microwave electronics,\cite{TvD81,Shen94,Lan97,Nis09} high-frequency superconducting materials issues,\cite{Hein99} and early microwave measurements of superconductors \cite{Oates11}.\\
	
	This article begins with a review of the basic features that distinguish superconductors from ordinary metals.  It then summarizes the key technical quantities that characterize the superconducting state, at least as far as microwave applications are concerned.  The heart of the paper is a discussion of major applications of superconductors in the microwave domain, and a discussion of the history of microwave superconductivity, emphasizing the main thrusts of applications. The article concludes with some forward-looking statements about possible future directions for this exciting field at the interface between science and technology.\\
		
\section{INTRODUCTION TO THE ESSENTIALS OF SUPERCONDUCTIVITY}
All superconductors are characterized by three universal hallmarks, namely zero DC resistance, the Meissner effect (think of floating magnets), and macroscopic quantum phenomena (quantum mechanics visible to the eye!).  

\subsection{Zero Resistance}
The zero resistance state of metals was discovered by H. Kamerlingh Onnes in 1911 (Nobel Prize in Physics in 1913). Onnes was the first to liquefy Helium, and found that the resistance of Mercury went to zero below a temperature of 4.2 K \cite{Delft10}.  The temperature at which DC resistance goes to zero in the limit of zero current is defined as the critical temperature, $T_c$.  This temperature is material specific.  Experimental values of $T_c$ range from 0.3 mK for Rh, to 9.2 K for Nb, to more than 30 K for $La_{2-x}Sr_xCuO_4$, to more than 145 K for the Hg-Ba-Ca-Cu-O family of cuprate superconductors, and is even approaching room temperature for a family of super-hydride materials, although they are stable only under extremely high pressure \cite{Droz15}.  These latter three families of compounds are examples of High-$T_c$ Superconductors (HTS), and have all been discovered since 1986.  A large number of materials have been found to be superconducting at low temperature, making superconductivity the preferred ground state for most metals \cite{Ham15}.  One interesting observation is that 'good' metals (e.g. low resistivity metals like Cu, Ag, Au) tend to be ‘bad’ superconductors (i.e. no measurable $T_c$), whereas ‘bad’ metals (high resistivity) tend to be ‘good’ superconductors (i.e. higher $T_c$ values or other useful superconducting properties).  This surprising correlation exists because often the mechanism that causes scattering in the normal state is also the mechanism that produces pairing of electrons in the superconducting state.  \\

	The most dramatic demonstration of zero DC resistance comes from measurements of persistent currents in closed superconducting rings \cite{Delft10}.  The circulating current creates a solenoidal magnetic field and the zero resistance state can also be used to generate very large and stable magnetic fields by making a superconducting solenoid \cite{Wilson12}.  Both magnetic resonance imaging and high-resolution nuclear magnetic resonance spectrometers are enabled by superconducting magnets  \cite{AnlageNMR01}.\\
	
	In terms of finite frequency properties, the superconducting state is characterized by the creation of an energy gap $\Delta$ in the electronic excitation spectrum.  A full energy gap over the entire Fermi surface turns a superconductor, somewhat paradoxically, into an insulator in the limit of zero temperature, at least for photons with energy less than the minimum value of the energy gap, $hf<2\Delta$ where $h$ is Planck’s constant and $f$ is the frequency of the radiation.  Hence such a superconductor can show a nearly zero loss microwave behavior in the limit of very low temperature.  This creates conditions for very high-Q resonators, and sets the stage for exploration of dramatic quantum effects, as we discuss below.\\
	
\subsection{The Meissner Effect}  
\begin{figure}
\centerline{\includegraphics[width=3.5in]{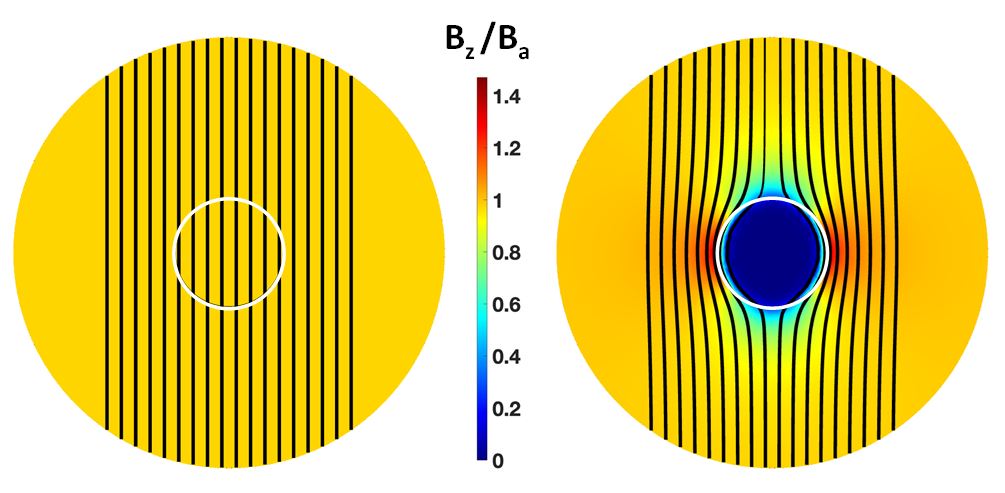}}
\caption{Illustration of the Meissner effect for a superconducting sphere (shown in cross section as a white circle).  Above the transition temperature a static externally applied magnetic field $B_a$ in the vertical direction uniformly permeates the normal metal sphere (left, yellow corresponds to $B_z=B_a$).  Below the transition temperature the superconductor spontaneously excludes the magnetic field (right, colors show $B_z/B_a$).  Shown are vertical cross sectional views through the center of the sphere.  The colors represent the magnitude of the z-component of magnetic field while the black lines are streamlines of the magnetic field.  The diameter of the superconducting sphere is 30 penetration depths.  The calculation is performed using time-dependent Ginzburg-Landau theory \cite{Ori20}. \label{MeissnerFig}}
\end{figure}

A superconductor can be distinguished from a mere perfect conductor (i.e. a metal with zero DC resistance) through the Meissner effect.  Consider a superconducting sample at a temperature above $T_c$ in a static external magnetic field, as shown in Fig. \ref{MeissnerFig}.  After some time, the eddy currents in the sample will have died away because of the sample’s finite resistance, and the magnetic field will be homogeneously distributed inside the sample.  If the material is now cooled below $T_c$, it will spontaneously develop screening currents which will actively exclude magnetic flux from the interior of the sample.  The result is shown in the right side of Fig. \ref{MeissnerFig} for a superconducting sphere.  Note that a material which transitioned from ordinary conductor to perfect conductor at $T_c$ would not show the Meissner effect in a static magnetic field.  It would instead trap the magnetic flux inside itself, as it became a perfect conductor.  The Meissner effect is unique to superconductors and arises from the quantum correlations created between electrons in the superconducting state.\\

The Meissner effect is best defined as the development of a (near) perfect diamagnetic state in a static external magnetic field, and is at the root of the magnetic levitation effect.  The Meissner effect demonstrates that superconductivity and magnetism are generally (although not universally) incompatible.  It implies that a large enough magnetic field applied to the sample can destroy superconductivity. \\

\subsection{Macroscopic Quantum Phenomena}  
The superconducting state is fundamentally and uniquely a quantum state of matter.  In other words it cannot be understood based entirely upon classical concepts.  For example, a single complex quantum wavefunction, which is phase coherent over macroscopic distances, can be used to describe the superconductor in many (but not all) circumstances.  This wavefunction describes a condensate of Cooper-paired electrons.  In the Ginzburg-Landau approach, the superconducting state can be described by a complex order parameter $\Psi(\vec{r}) = |\Psi(\vec{r})| e^{i\phi(\vec{r})}$, where $\phi(\vec{r})$ is the position-$(\vec{r})$ dependent phase factor.  As such, the material can show unique macroscopic quantum phenomena such as the Josephson effect, magnetic flux quantization, and microscopic quantum superposition states and entanglement.\\

The order parameter must be single-valued throughout the superconductor.  This in turn implies that $\phi(\vec{r})$ returns to the same value (modulo $2\pi$) for any closed circuit taken through the superconductor.  Consider a superconductor which incorporates a hole (i.e. a doughnut), or containing a finite bounded region in which the order parameter $|\Psi(r)|\rightarrow 0$.  Following a path $C$ through this material, which encloses the hole, will lead to the conclusion that the magnetic flux $\Phi = \int_{C}  \vec{A} \cdot d\vec{l}$ must be quantized in integer multiples of the quantum of magnetic flux $\Phi_0 = h/2e$.  Here the line integral is over the dot product of the vector potential $\vec{A}$ along a closed circuit $C$ that lies entirely inside the superconductor.  This unit of flux involves only fundamental constants of nature (Planck’s constant and the charge of the electron) and the factor of 2 arises from the microscopic phenomenon of Cooper pairing of the charge carriers in the metal.  Flux quantization imposes important constraints on any closed-loop superconducting circuit, and when combined with the Josephson effects leads to surprising new phenomena.\\

Brian Josephson predicted that pairs of electrons could tunnel through a classically forbidden region (barrier) between two superconductors even at zero potential difference \cite{Jos62}.  Consider two superconducting banks, each described by a macroscopic quantum wavefunction with independent phases $\phi_1$ and $\phi_2$, separated by a thin insulating barrier.  The DC pair tunnel current through the barrier is given by $I=I_{cj} \sin \delta$, where the gauge-invariant phase difference $\delta=\phi_1-\phi_2-\frac{2\pi}{\Phi_0}\int \vec{A}\cdot d\vec{l}$ includes the effects of magnetic field in the junction (through the vector potential $\vec{A}$), and $I_{cj}$ is the critical (or maximum) current of the junction. This predicts that a spontaneous supercurrent will flow between the two superconductors, and its magnitude and direction can be controlled through electromagnetic means.   Josephson also predicted that a voltage difference $V$ imposed between the two superconducting electrodes will cause the phase difference to increase linearly with time $t$ as $\Delta \phi = 2e V t/\hbar$, where $\hbar$ is Planck’s constant $h$ divided by $2\pi$.  Putting this back into the first Josephson equation results in a supercurrent between the electrodes which oscillates at frequency $\omega = 2eV/ \hbar$.  Note that the imposed voltage and resulting frequency of oscillation of the Cooper-pair tunnel current are directly related by means of universal constants of nature.  These two simple Josephson effects have given rise to many remarkable microwave applications, including new computational paradigms that are poised to revolutionize our digital computing technology.  Josephson won the Nobel Prize in Physics for this work in 1973 (along with Esaki and Giaever).\\

The Bardeen-Cooper-Schrieffer (BCS) theory of superconductivity \cite{BCS} has at its heart a ground state superconducting wavefunction that includes the subtle quantum correlations between all of the charge carriers in the metal.  The theory (along with its many generalizations) is essentially exact, at least for a broad class of superconducting materials, and forms a very firm foundation for superconducting microwave technology.  This work earned the authors the Nobel Prize in Physics in 1972.  Next we discuss the essential phenomenology of superconductivity that is relevant for understanding microwave applications of superconductors.\\

\section{PHENOMENOLOGY OF SUPERCONDUCTIVITY ($T_c$, $H_c$, $J_c$, $\omega_c$)}
\label{SecIII}
There are strict limits to the domain of superconductivity.  Superconductivity is destroyed for temperatures above $T_c$ because the thermal agitation destroys the subtle quantum correlations between electrons that constitute the superconducting state.  Due to the general incompatibility of magnetism and superconductivity, there is a limit to how large a magnetic field a superconductor can exclude in the Meissner state.  This is characterized by the critical field, $H_c$.  An estimate of the critical field comes from comparing the energy density of the magnetic field required to destroy superconductivity to the free energy gain of the superconducting state:
$\mu_0 H_{c}^{2}/2 = f_n(T)-f_s(T)$ where $f_n$ and $f_s$ are the Helmholtz free energy densities in the normal and superconducting state at temperature $T$ and zero field.  This thermodynamic critical field $\mu_0 H_{c}$ can exceed 1 Tesla at low temperatures, depending on the material.  Related to this, the superconductor is able to support large zero frequency current densities, $J$.  These currents carry significant kinetic energy because the currents flow without dissipation or scattering.  The critical current density, $J_c$, is reached when the kinetic energy in the current carried by the superconductor equals the free energy gain of the superconducting state over the normal state.  Silsbee’s rule states that when the surface self-magnetic field created by the current in a round conducting wire approaches the critical field, superconductivity will be destroyed.  The corresponding critical current density can often exceed $10^9$ A/m$^2$.\\

Finally there is a frequency limit to superconductivity due to the finite binding energy of the Cooper pairs that make up the superconducting condensate.  The gap frequency $\omega_c = 2\Delta/\hbar$ corresponds to the photon energy that directly breaks Cooper pairs into un-paired quasiparticles, thus degrading the superconductor.  The range of gap frequencies $f_c=\omega_c/2\pi$ vary from about 20 GHz for some low-$T_c$ superconductors to the THz range for high-$T_c$ cuprate superconductors.  Hence for frequencies substantially above $\omega_c$ the superconductor basically responds the same way as it would in the normal state.  Thus superconductors have infrared and visible wavelength properties that are essentially no different when compared just below and just above the superconducting transition temperature $T_c$.\\ 

Superconductors come in two flavors, Type I and Type II.  They are distinguished by their response to a magnetic field.  A Type I superconductor usually does not compromise with the magnetic field; it is either superconducting in the Meissner state, or it is a normal conductor when the applied magnetic field exceeds the thermodynamic critical field, $H_c$.  However, depending on the geometry of the sample, a type-I superconductor can enter a “compromise state” known as the intermediate state.  The regions of superconducting and normal material act in some sense like two immiscible fluids because the superconductor/normal interfaces are energetically costly.  Type II superconductors, on the other hand, will compromise with the magnetic field and create a “mixed state” in which magnetic field is allowed to enter the superconductor but only in discrete flux-quantized bundles, called magnetic vortices.  In this case the superconductor/normal interface is energetically favorable, which results in a proliferation of the interfaces such that each vortex carries an integer number of magnetic flux quanta, $\Phi_0$, ultimately as few as one.  Microwave measurements were among the first to clearly demonstrate that magnetic vortices exist in type-II superconductors.  The measurements of Gittleman and Rosenblum demonstrated that vortices act as coherent entities that experience a Lorentz-like force in the presence of an alternating current, and encounter both pinning and viscous drag forces as well \cite{Git66}.\\

\subsection{Microwave Screening Properties} 
	Our main concern here is the response of superconductors to high frequency electromagnetic fields.  Most ordinary metals have a conductivity $\sigma$ that is real and frequency independent in the microwave to mm-wave frequency range.  Solving Maxwell’s equations for a plane wave impinging on a normal metal satisfying the Ohm’s law local constitutive equation ($\vec{J}=\sigma \vec{E}$) results in a complex surface impedance given by $Z_s=(1+i)/\sigma \delta$, where $\delta(\omega)=\sqrt{2/\mu_0 \omega \sigma}$ is the frequency dependent skin depth.  The tangential electric field obeys $E(z) \sim e^{-ikz}$ with complex wavenumber $k = \sqrt{i\omega \mu_0/Z_s}$ as a function of depth $z$.  This reveals that microwave currents will flow both in-phase and in quadrature to the imposed electric field, and these currents will both oscillate and decay as a function of depth into the material as $J_n(z)\propto e^{iz/\delta(\omega)}  e^{-z/\delta(\omega)}$.  A significant limitation of normal metals is their strong dispersion in the oscillation and decay characteristics as a function of frequency, known as the 'skin effect'.  This handicaps the ability of normal metal transmission lines to carry broadband information \cite{Kautz78,Mat95}.\\
	
	\begin{figure}
\centerline{\includegraphics[width=2.0in]{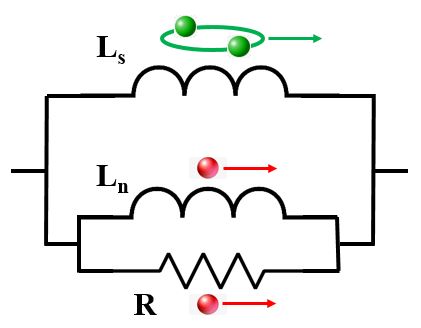}}
\caption{Schematic of the two fluid model of superconductor response
to microwave fields. A finite frequency electric or magnetic field will
induce currents in both the superfluid (upper branch) and the normal fluid (lower branch). The superfluid channel is purely inductive ($L_s$) in nature and the current is carried by Cooper pairs. The normal fluid channel has both resistance $R$ and (usually small) inductance ($L_n$) and is carried by the normal fluid. The complex impedance of this simple circuit gives an excellent qualitative picture of superconductor response as a function of frequency and temperature. \label{Two_Fluid_Fig}}
\end{figure}

	Superconductors on the other hand have a complex conductivity ($\sigma=\sigma_1-i\sigma_2$) that is primarily imaginary ($\sigma_2 \gg \sigma_1$), giving rise to a surface impedance that is dominantly reactive, $Z_s=R_s+iX_s$ with $X_s>>R_s$ at low temperatures.  The resulting screening currents in the superconductor (required by the Meissner effect) show a simple frequency-independent penetration depth, $\lambda$, as $J_s(z)\propto e^{-z/\lambda}$, with no oscillation.  The screening properties are frequency independent up to frequencies on the order of the superconducting energy gap, $\Delta/h$, which can be in the THz range.  For future reference, note that in the low frequency limit (below the critical frequency $\omega_c$) $\sigma_2=1/(X_s \lambda)=1/(\mu_0 \omega \lambda^2 )$.\\
	
		How does a superconductor respond to an imposed electric field tangent to its surface?  This can be approximately described by London’s phenomenological equations.  The first states that the superconducting electrons will be accelerated by the electric field, with no dissipation due to scattering: $\partial (\Lambda \vec{J}_s)/\partial t=\vec{E}$, where $\Lambda=m/(n_s (e)^2 )=\mu_0 \lambda_L^2$ and $\vec{J}_s$ is the supercurrent density.  We can think of the London penetration depth $\lambda_L$ as the ideal magnetic penetration depth $\lambda$ when all of the electrons in the metal participate in the Meissner screening.  The first London equation says that in order to create an alternating current (i.e. any current at a non-zero frequency) it is necessary to establish an electric field in the superconductor.  However, this has implications for the finite-frequency losses in superconductors.  If any un-paired electrons (quasiparticles) are present, they will be accelerated by this electric field, scatter, and cause Ohmic dissipation.  The second London equation defines the response of a superconductor to a magnetic field, $\vec{\nabla} \times \vec{J}_s=-\vec{B}$, which holds for both static and dynamic fields.  Hence this equation shows that a magnetic field can be used to induce screening supercurrents (DC or AC), in a dual manner to how a current is induced in a normal metal with an electric field. \\
		

	To gain a good qualitative understanding of the microwave properties of superconductors one can consider the two-fluid model of loss and inductance (see Fig. \ref{Two_Fluid_Fig}).  The model is very simple, but contains a number of key features that are consistent with experiment, and qualitatively in agreement with full microscopic theory \cite{Bar2F}.  A superconductor is thought to have two independent fluids, one made up of superconducting electrons, the other of normal electrons, that inter-penetrate and act in parallel, but do not interact.  The relative abundance of these two fluids changes as a function of temperature.  We say that the superfluid has a number per unit volume of $n_s(T)$, while the normal fluid has a number density of $n_n(T)$.  The total number density is equal to that of the metal in the normal state: $n_s(T) + n_n(T) = n$, a number density fixed by the nature of the metal.\\  
	
	The total conductivity of the superconductor is the sum of the superfluid and normal fluid components: $\sigma=\sigma_s+\sigma_n$.  There is a simple circuit analogy that captures the essential features of this complex conductivity (Fig. \ref{Two_Fluid_Fig}).  The superconductor acts as if it is a parallel connection of a resistor $R$ (representing the normal channel) and a pure lossless inductor $L_s$ (representing the superfluid channel).  At zero frequency all of the current goes through the inductor, and there is no Ohmic loss (the property of infinite DC conductivity).  At finite frequency the inductive channel now presents some non-zero impedance ($Z_{super} = i \omega L_s$) and as a result some of the current is shunted into the resistive channel (we ignore the usually small $L_n$).  The relative population of the normal and super channels depends on frequency and temperature as  $J_s/J_n = \sigma_{2s}/\sigma_{1n} = \frac{n_s}{n_n} \frac{1}{\omega \tau_{n}}$, where $\frac{1}{\tau_{n}}$ is the scattering rate of the normal fluid electrons.  Since it is often the case that $\omega \tau_{n}<<1$ the ratio $J_s/J_n$ is usually much larger than 1, meaning that most of the current flows through the super-channel until one reaches frequencies near the superconducting gap frequency $\omega_c$, or near the transition temperature where $n_s(T)$ is very small.
	
	It is important to understand the frequency dependence of the dissipated power in a superconductor due to a current density $J$.  The dissipated power per unit volume can be calculated from  $P=Re[\rho] J^2 = Re[1/\sigma] J^2$.  This results in  $P=\frac{\sigma_1}{\sigma_1^2 + \sigma_2^2} J^2$.  For a superconductor at “low frequencies” such that $\omega \tau_{n}<<1$, we can take  $P_s \approx \frac{\sigma_1}{\sigma_2^2} J^2$.  To good approximation we can take $\sigma_1$ to be independent of frequency and we know that $\sigma_2 \propto 1/\omega$ from above, hence for a superconductor we expect $P_s \propto \omega^2$.  The corresponding calculation for a normal metal results in a dissipated power per unit volume $P_n \propto \omega^{0}$, but the total dissipated power scales as $P_{n,Total} \propto \omega^{1/2}$ because the skin depth scales as $1/\omega^{1/2}$.  Hence superconductors start with far smaller loss than normal metals at low frequency, but Ohmic losses can ultimately exceed normal metals beyond a crossover frequency, typically at frequencies above 100 GHz for cuprate superconductors \cite{Nis09}.\\

\section{MICROWAVE TECHNOLOGIES ENABLED BY SUPERCONDUCTIVITY}
Here we introduce the high frequency applications of superconductors that follows from each unique property discussed above.

\subsection{Low-loss properties at microwave frequencies} 
As discussed above, superconductors only display zero dissipation at precisely zero frequency.  All finite frequency electromagnetic stimulations of a superconductor result is some (usually very) small dissipation.  Heinz London was the first to show experimentally that superconductors have non-zero microwave loss below the transition temperature,\cite{London40} and Pippard later showed that superconductors present considerable microwave reactance as well \cite{Pippard53}.  Numerous microwave applications of superconductors rely on the low microwave losses present in the superconducting state.    Microscopically, losses come from un-paired electrons that directly absorb microwave photons, as described originally in the microscopic theory of Mattis and Bardeen \cite{MB58}.  For a fully-gapped superconductor the losses are exponentially small in the limit of zero temperature because creating unpaired electrons requires that they be thermally activated over a finite energy gap $\Delta$ \cite{Turn91}. The microwave surface resistance $R_s$ of a superconductor, which is proportional to the dissipated power $P$, is given by $R_s \propto P \propto \frac{(\hbar \omega)^2}{k_B T} ln(\frac{4k_B T}{\hbar \omega}) e^{-\Delta/k_B T}$, valid for $T/T_c < 1/2$, $\hbar \omega \ll \Delta$, and $\hbar \omega \ll k_B T$.  This low-loss limit enables highly efficient microwave particle accelerator cavities with quality factors $Q$ exceeding $10^{11}$  \cite{Pad98}.  The cavities are designed to convert microwave energy into the kinetic energy of a charged particle beam.  The objective is to maintain the low-loss properties (high $Q$) up to high accelerating gradients (large RF electric field on the accelerating axis of the microwave cavity) to create compact and efficient accelerator structures.  To excel at this application, the superconducting material must satisfy many constraints, and so far only a few materials have shown promise.  Tremendous progress has been made in optimizing the surface properties of bulk Nb used in these cavities.  Other superconducting materials are also under development, such as Nb films on copper substrates, and Nb$_3$Sn coatings \cite{Posen15}.  It is expected that compact superconducting accelerator structures will find wider applications as their efficiency increases and as materials with higher transition temperatures are successfully utilized in high-Q cavities. \\    

Another application enabled by low microwave losses are high-performance microwave band pass filters \cite{Ham11}.  The high Q values of the individual resonators representing poles of the filters allow for design of extremely selective bandpass filters, as illustrated for example in Refs. \cite{Will01,Berk02,Nis09}.  Superconducting filter structures can be made physically smaller than their normal metal counterparts because of their superior current-handling capabilities and the high dielectric constant substrates used for growing HTS films.  These properties allow two-dimensional superconducting structures that are as good as, or superior to, three dimensional normal metal and dielectric filter structures, effectively reducing the dimensionality of the structures and saving space and weight.  The filters have extremely low insertion loss in the pass band, and the cryogenic environment allows for inclusion of low-noise amplifiers directly behind the filters, capitalizing on the low-noise environment of the cryo-platform.\\

\begin{figure}
\centerline{\includegraphics[width=3.5in]{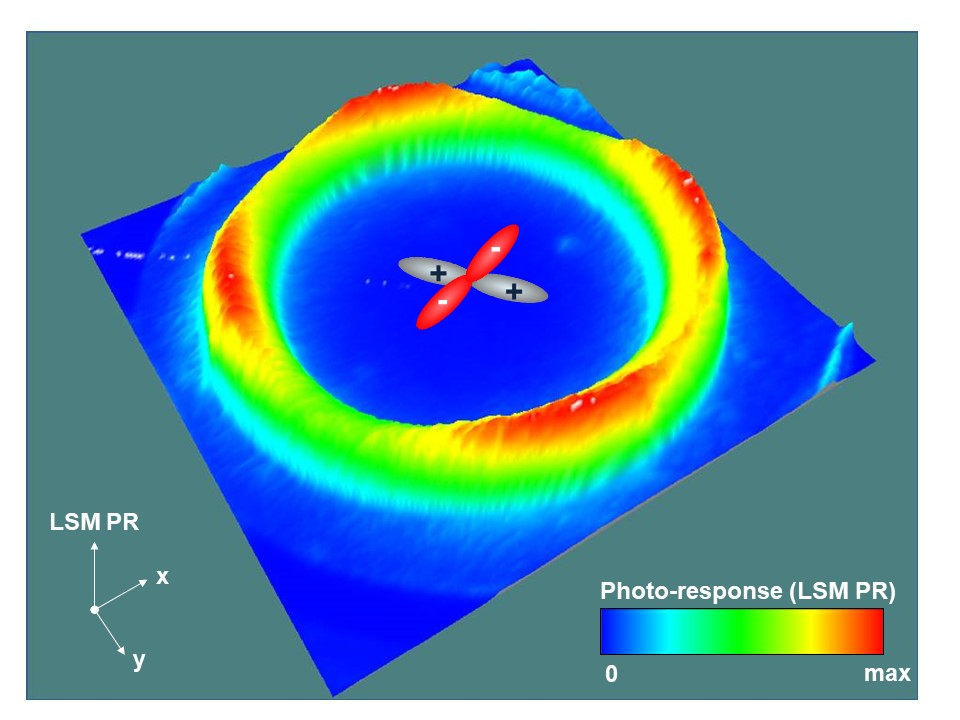}}
\caption{Photoresponse (PR) image of a superconducting spiral resonator standing wave pattern at 256.67 MHz and 3.039 K.  The PR is a convolution of the RF current standing wave pattern and the anisotropic nonlinear Meissner effect of this nodal superconductor.  The resonator is a thin film of $YBa_2Cu_3O_{7-\delta}$ on an $MgO$ substrate with well-defined crystallographic orientation.  The grey and red lobes at the center show the superconducting order parameter of the film, as deduced from the angular dependence of the photoresponse.  For more details see Refs.\cite{Zhu13,Zhu18}.  Image courtesy of Dr. A. P. Zhuravel of the B. I. Verkin Institute for Low Temperature Physics and Engineering in Kharkov, Ukraine. \label{LSMPR}}
\end{figure}

The low-loss properties of superconductors are limited by high power microwave signals that induce currents approaching the critical current of the superconducting components \cite{Zhu02}.   Before that however, superconductors harbor a number of nonlinearities, both intrinsic and extrinsic, and these can limit the performance of superconducting microwave devices \cite{OatesNL01}.  In addition, if the superconductor has nodes in its energy gap, such as d-wave superconductors (e.g. most high-T$_c$ cuprates), then a number of qualitatively new phenomena appear \cite{Mao95}.  First is the existence of enhanced nonlinearity at low temperatures associated with the quasiparticle excitations near the nodes of the superconducting gap.  This gives rise to enhanced intermodulation distortion,\cite{Oates04} and a remarkable anisotropic nonlinear Meissner effects of both diamagnetic \cite{Zhu13} and paramagnetic character \cite{Zhu18}.  Figure \ref{LSMPR} illustrates the diamagnetic anisotropic nonlinear Meissner effect in an HTS RF resonator, imaged with a cryogenic laser scanning microscope \cite{Zhu06}.  In addition, the low temperature losses have an intrinsic finite residual conductivity $\sigma_{min} = ne^2/(\pi m \Delta_0)$, where $\Delta_0$ is the maximum value of the gap on the Fermi surface \cite{Hirs93}.  Also there is a linear temperature dependence of the surface resistance in the clean limit, $R_s(T) \sim T$ at low temperatures.  Hence nodal superconductors are not employed for ultra-low loss applications, and must find their niche elsewhere.\\

\subsection{Near-Zero Dispersion of SC Transmission Lines} 
	The lack of dispersion in superconducting transmission lines makes them very attractive for high-speed electronics,\cite{Kautz78,Ek90,Mat95} and forms the basis for multiple generations of superconducting electronics and digital computing, beginning in the 1970’s \cite{Ana80}.  Josephson junctions can undergo a rapid $2\pi$ phase slip, as fast as time scales on the order of $h/\Delta \sim $ few ps for Nb.  The resulting quantized voltage pulse $V(t)$ has the property that $\int V(t)dt=\Phi_0$, which is the flux quantum, introduced earlier.  The presence or absence of such voltage pulses can act as digital bits, and a decision-making logic can be constructed with other Josephson junctions. Several generations of such Josephson-based high-speed digital superconducting logic have been developed \cite{Kot90}.  \\
	
	Low-loss superconducting transmission lines also offer the ability to support slow waves and to create compact delay lines, as well as other types of analog microwave devices \cite{With89,OatesAn01}.  The large kinetic inductance of superconductors (see below) is strongly dependent on temperature and DC transport current \cite{Mes69} allowing for widely variable microwave delay lines and amplifiers \cite{Eom12,Zhao17}.\\

\subsection{Superconducting Kinetic Inductance}
	
	Kinetic inductance arises from the inertia of the current-carrying charge carriers, and acts in series with the magnetic inductance of a conductor.  Heinrich Hertz had set out to measure the inertia of charge carriers at the age of 21 \cite{Hertz96}.  He was motivated to work on this topic when a notice for a Prize was posted at the Friedrich-Wilhelms University in Berlin in the Fall of 1878.  The Prize was for a solution of the problem of electrical inertia, and he was given access to a laboratory by Prof. von Helmholtz to perform the experiments.  His first scientific publication, “Experiments to Determine and Upper Limit to the Kinetic Energy of an Electric Current,” was devoted to this problem.  Hertz had attempted to measure this form of inductance, but found that it was at least 300 times smaller than the geometrical inductance, concluding that it was not relevant to the electrical properties of ordinary conductors  \cite{Hertz96}.  The first explicit experimental demonstration of superconducting kinetic inductance was apparently made by W. A. Little in 1967 \cite{Little}.  Soon after, Meservey and Tedrow measured the kinetic inductance of thin Tin (Sn) films and noted that the kinetic inductance is extremely sensitive to temperature changes, magnetic field, and current \cite{Mes69}.\\
	
	Superconducting kinetic inductance is directly related to the superfluid density $n_s$ as $L_{kinetic}=m/(n_s e^2 ) (L/A)$ for a superconducting wire carrying a uniform current along its length $L$ and cross-sectional area $A$.  Thus any disturbance that reduces the superfluid density will result in enhanced kinetic inductance.  Examples include temperatures near $T_c$, large transport currents approaching the critical current density $J_c$, and large magnetic fields approaching $H_c$.  However the reduction of superfluid density is accompanied by the creation of quasiparticles, and a corresponding increase in surface resistance and dissipated power.  Hence obtaining large kinetic inductance in this manner is most suitable for low frequency applications such as very precise thermometry,\cite{McD87} unless Josephson inductance is employed (see below).\\
	
	A new class of radiation detectors have been built based on the sensitivity of superconducting kinetic inductance to temperature changes.  In this case the inductor is part of a resonant microwave circuit, so that changes in kinetic inductance translate into frequency shifts of the resonating element  \cite{Day03}.  The ability to create compact superconducting resonant circuits (see below) using lumped-element components allows many such resonators to be in a focal plane array of an imaging system.  Each resonator can be tuned to a distinct frequency such that large numbers of them can be monitored by means of a single interrogation transmission line  \cite{Z12}.  In order to minimize losses from absorption of radiation and the creation of quasiparticles, granular superconducting materials are often employed in the inductor.  These materials have small superconducting grains that are Josephson-coupled to each other, enhancing their inductance and mitigating their loss due to creation of quasiparticles to some extent.  These microwave kinetic inductance detectors (MKIDs) have proven to be very useful for imaging weak sources of electromagnetic radiation over a broad range of frequencies above $\omega_c$. \\
		
	In fact, the Josephson inductance of engineered junctions offers an alternative tunable inductance that can be accompanied with low microwave losses.  The inductive response of a Josephson junction is characterized by an effective Josephson inductance  $L_{JJ}=\frac{\Phi_0}{2 \pi I_{cj}\cos \delta}$, where $\delta$ is the gauge-invariant phase across the junction, including the magnetic field in the barrier through the vector potential $\vec{A}$  \cite{Feld75,Rif76,Orl91}.  The nonlinear Josephson inductance has been used for many years to directly detect radiation through a variety of means  \cite{Rich73}.  By incorporating the junction into a superconducting loop one can manipulate $\delta$ by means of the magnetic flux threading the loop.   Such a structure (originally known as an RF superconducting quantum interference device - RF SQUID \cite{Silver67}) acts as a flux-tunable resonant circuit with strongly tunable and nonlinear properties.  It has been used to add variable inductance to superconducting transmission lines,\cite{Luine99} and as a meta-atom for new forms of artificial matter \cite{DCL06,Laz07,Jung13,Trep13,Laz18}.  Josephson inductance, and it's nonlinearity, are a key ingredient for superconducting quantum circuits \cite{Krantz19}.\\
	
\subsection{Compact Superconducting Structures}
 A hidden advantage of superconductors over normal metal components is the fact that superconductors can support much larger current densities but still maintain low losses.  This enables very compact structures to be built that survive under high current densities without significantly degrading their superconducting properties.  Hollow metallic resonators have quality factors that scale as $Q \propto V/S\delta$, where $V$ and $S$ are the volume and surface area of the resonator and $\delta$ is the depth of penetration of the electromagnetic fields in the metals making up the walls of the resonator.  In general one finds high-Q values in three-dimensional structures where $V/S$ can be made large.  However, because of their unique and low dissipation properties, and the ability to support large current densities, superconductors enable high-Q planar (quasi-two-dimensional) structures.  This has led to the development of compact highly-selective and low insertion loss planar bandpass filters.  It also facilitates extreme sub-wavelength meta-atoms to create effective media metamaterials with precisely defined and low-loss properties.  Example meta-atoms include split-ring resonators \cite{Ricci05} and compact spiral resonators \cite{Kurter10} that are as small compared to the resonant microwave wavelength as a Hydrogen atom is to visible light.

\subsection{\textit{Macroscopic} Quantum Phenomena}
	The superconducting state can only be understood microscopically through its quantum mechanical properties.  A phase coherent many-particle quantum wavefunction governs the ground and excited states of a superconductor.  Several remarkable macroscopic quantum phenomena follow from this property.  As introduced in Section~\ref{SecIII}, a vortex is a continuous strand of suppressed superconducting order parameter, upon which the magnetic field is centered, that can reach deep into a superconductor.  It contains a unit of magnetic flux and associated screening current, and can have elastic properties, interactions with other vortices, and experience forces from pinning sites or structures.  This entity in some sense enjoys a life of its own, and essentially uses the superconductor simply as a medium in which to exist.  In response to microwave currents the magnetic vortex will experience an oscillating Lorentz force and produce both reactive and dissipative response \cite{Git66,Coff91}.  Vortices are often deemed undesirable for many microwave applications.  For example they can produce residual loss in SRF accelerator cavities, and they can disrupt high-speed Josephson based digital computing circuits.  Ideally vortices are either eliminated from the material, or are relegated to “moats” \cite{Ber83,Sem16} that effectively immobilize them and sequester them from microwave currents.\\

High-speed superconducting digital logic is based on the use of a single magnetic flux quantum as the classical bit.  A Josephson junction can be sent through a $2\pi$ phase winding of its gauge-invariant phase on time scales as short as the inverse gap frequency $\tau = \hbar/\Delta$, which is on the order of a few ps for Nb.  The time-dependent phase difference creates a voltage pulse $V(t) \propto \frac{d\delta}{dt}$ which is on the scale of $\Delta/e$ in magnitude and duration on the scale of $\tau$.  This voltage pulse has the property that it contains a unit of magnetic flux: $\int V(t) dt = \Phi_0$.  The presence or absence of such a pulse at logic gates (made up of other Josephson junctions) acts as the classical bit for logic operations.  The use of low-dispersion and low-loss superconducting transmission lines (as discussed in Sections IV A. and B. above) preserves the integrity of these pulses as they propagate through the logic circuits.  This logic scheme was dubbed the rapid single-flux-quantum (RSFQ) circuit family,\cite{Lik91} and has matured into a number of derivative logic families of high speed and low power consumption.  An RSFQ-based digital frequency divider has been operated up to 750 GHz \cite{Chen98}.  It has become clear that low-dissipation-per-operation logic is absolutely critical for moving beyond peta-flop computing,\cite{Hol13,Man15} and a number of superconducting logic families show promise in this regard \cite{Her11,Osb19}.    Designing and building these advanced superconducting digital technologies poses many exciting challenges in microwave engineering \cite{Br06}.\\

\subsection{\textit{Microscopic} Quantum Phenomena}
As noted in Section II C. above, the superconducting state is defined by a remarkable macroscopically coherent many-electron quantum wavefunction. Under the right conditions this property enables superconductors to display \textit{microscopic} quantum phenomena even when the devices involved are
macroscopic in size. Think about the famous ‘particle in a box’ one-dimensional quantum mechanics problem covered in undergraduate textbooks \cite{Grif05}. The quantum energy states of the particle are discrete and widely separated, labeled by positive integers, and the corresponding wavefunctions are sinusoidal patterns that span the length of the box. In the case of a superconducting quantum bit (qubit) the particle is a bit more abstract, being essentially a ‘phase point’ in a ‘box’ described by the potential energy of a Josephson junction. This device is a superconducting system utilizing one or more Josephson junctions with two nearby energy levels, a ground state $|g>$ with energy $E_g$ and a first excited state $|e>$ with energy $E_e$, with all other higher energy levels safely separated such that the system can be maintained in this limited (two-state Hilbert space) manifold. The energy difference between the two states of the qubit is chosen to be in the microwave range, $f_{qubit}=(E_e - E_g)/h$,
with $f_{qubit} \approx 5$ GHz. Hence the quantum state of the qubit can be manipulated by means of microwave photons.  To exhibit quantum effects, the temperature of the qubit must be maintained below the point at which thermal excitations cause transitions between the two states, namely $k_B T \ll h f_{qubit}$. Since a 5 GHz photon has an equivalent thermal energy corresponding to 240 mK, superconducting qubits are typically operated at temperatures below 20 mK.  This may sound like an extreme condition, but in fact such temperatures are routinely achieved with commercially available automated cryostats that are fully compatible with microwave transmission lines, amplifiers, circulators, etc.\\

The earliest superconducting qubits utilized higher transition frequencies to ease the cooling requirements \cite{Fried00}. Flux-based qubits are typically designed as compact self-resonant structures based on lumped-elements that also include one or more Josephson junctions incorporated into superconducting loops. The earliest versions were patterned on the basic structure of a Superconducting Quantum Interference (SQUID) device, which is just a superconducting loop interrupted by one, two, or more Josephson junctions. Because of the macroscopic quantum properties of flux quantization and the Josephson effect, an applied magnetic field (specifically the magnetic flux applied to the loop) can be used to control the gauge invariant phase difference $\delta$ on the Josephson junctions. This in turn allows one to control the Josephson potential energy landscape to create a quantum ‘particle in a box’ scenario that produces the requisite ground and first excited states
making up the qubit. The nonlinearity of the Josephson potential energy then facilitates the isolation of the two lowest energy states from the others during subsequent microwave signal manipulations.\\

The next step is to carefully apply microwave signals of precisely controlled frequency and duration to manipulate the quantum state of the qubit \cite{Rist20,Bard20}. These signals are created with arbitrary waveform generators with precisely controlled in-phase and quadrature content. The signals perform manipulation of the qubit state and can best be visualized in terms of a point representing the quantum state of the two-level system on the Bloch sphere \cite{Bard20}.  The microwave manipulations can be used to perform fundamental 'gate' operations, such as a $\pi$ pulse that inverts the state of the qubit, or various $\pi/2$ rotations that create non-trivial superposition states of $|g>$ and $|e>$ \cite{Krantz19}.  After demonstrating control over individual qubits, the creation of multi-qubit systems greatly complicates the microwave engineering issues.  To first approximation each qubit must be completely isolated from all the others, and no microwave cross-talk can occur for the control signals.  At the next level of operation, the qubits must be brought into precisely controlled interaction to spread the quantum information so that it eventually entangles all the qubits in the entire quantum processor.  The engineering details to accomplish these tasks are extremely intricate and demanding \cite{Rist20,Bard20,Rose20}.  \\

Many ancillary operations with superconducting qubits take advantage of the other microwave properties mentioned above.  For example it is important to control the electromagnetic impedance that the qubits experience at microwave frequencies \cite{Rose20}.  The quantum states are extremely delicate and prone to being lost due to electromagnetic interactions with the environment.  One issue is the presence of parasitic two-level systems in dielectrics that mimic the properties of the qubits and can couple to them and destroy their quantum information \cite{Oliv13,Mull19}.  More distant perturbations can be controlled by placing the qubit in a high impedance environment.  This can be accomplished by coupling it to large inductances (called superinductors), such as that provided by high kinetic inductance materials or by arrays of closely spaced Josephson junctions \cite{Krantz19}.  The scale for the impedance required is given by the quantum of resistance $R > R_Q =h/(2e)^2 \approx 6.45 k\Omega$\\

		\begin{figure*}
\centerline{\includegraphics[width=7.16in]{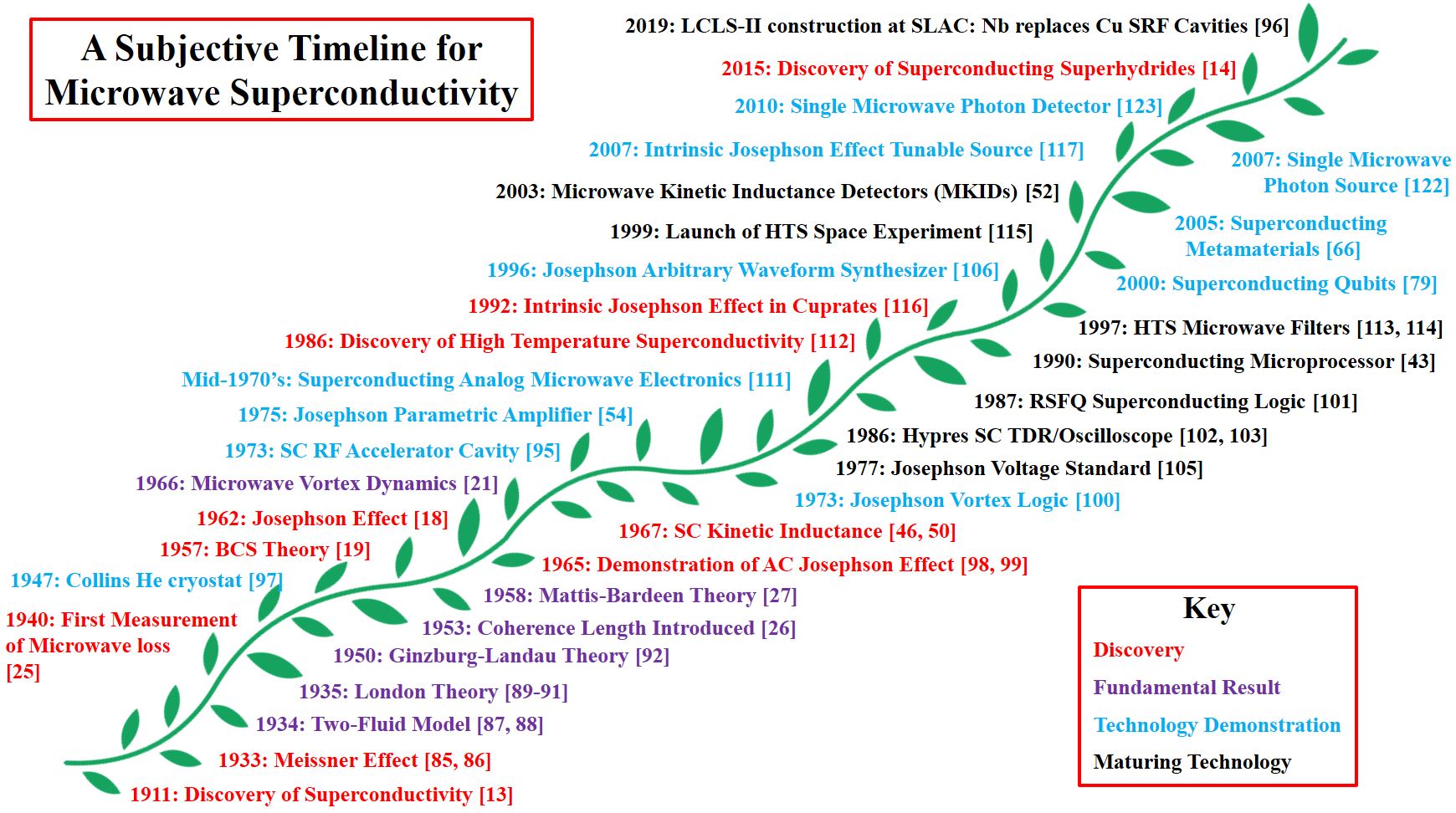}}
\caption{The author's subjective timeline of major events in the history of microwave superconductivity, from the discovery of superconductivity to circa 2020.  Events are arranged in chronological order and (the believed) seminal references are given. The events are color coded as Discovery (red), Fundamental Result (purple), Technology Demonstration (blue), and Maturing Technology (black).\label{TimeLineFig}}
\end{figure*}

\section{A TIMELINE OF MICROWAVE SUPERCONDUCTIVITY}

Figure \ref{TimeLineFig} offers the author's personal perspective on the timeline of important events in the development of microwave superconductivity.  The foundation for these developments was laid by the discovery of superconductivity in the lab of K. Onnes in 1911 \cite{Delft10} (Nobel Prize in Physics in 1913).  The demonstration of the spontaneous magnetic flux-excluding properties by Meissner and Ochsenfeld in 1933 \cite{M33,Forr83} showed that superconductivity was a uniquely non-classical phenomenon and was intimately related to electromagnetism.   Our understanding of the electrodynamic properties of superconductors was developed through the phenomenological models, along with the early experimental work of London \cite{London40} and Pippard \cite{Pippard53} on the microwave properties of superconductors. The two-fluid model \cite{Gorter34,Gorter55} and the London theory \cite{London35a,London35b,London50} of superconductor electrodynamics, along with Ginzburg-Landau theory,\cite{Ginz09} provided a phenomenological understanding of how electromagnetic fields interact with superconductors.  Landau won the Nobel Prize in Physics in 1962, while Ginzburg won in 2003 (along with Abrikosov and Leggett).  

Heinz London was the first to measure non-zero resistance below the transition temperature of a superconductor, and he attributed the measured microwave losses to residual normal fluid as proposed by the two-fluid model \cite{London40}.  Pippard was able to show through microwave surface impedance measurements that a new length scale was required to understand the penetration depth data taken from samples of varying purity, and this led to the concept of a coherence length, which proved to be an important ingredient for later microscopic theories of superconductivity \cite{Pippard53}. Microwave measurements were also instrumental in proving the existence of quantized magnetic vortices in type-II superconductors. The measurement of microwave loss vs. frequency of a superconductor in a magnetic field by Gittleman and Rosenblum was interpreted in terms of the coherent motion of vortex-particles subjected to a Lorentz-like force and hindered in their motion by pinning and viscous damping forces \cite{Git66}.  Their model has stood the test of time and is still considered a definitive starting point for treatments of microwave vortex response.  More sophisticated treatments of vortex motion under microwave stimulus now include the effects of flux creep,\cite{Coff91} and vortex elastic properties,\cite{Brandt91} among other effects \cite{Pomp08}.\\

The Bardeen-Cooper-Schrieffer (BCS) theory laid out a microscopic understanding of the superconducting state and showed that the many-particle quantum wavefunction that describes all of the electrons in the metal has a macroscopic phase-coherence and rigidity that explain many unique properties,\cite{BCS} such as the macroscopic quantum effects.  The essential quantum nature of the superconducting state has enabled many applications, and fueled the rise of quantum technology in the 21st century.  In particular the Mattis-Bardeen theory \cite{MB58} of superconductor electrodynamics demonstrated the importance of the superconducting energy gap and quantum coherence effects in the complex conductivity and surface impedance of superconductors.  Mattis-Bardeen theory predicts that losses in fully-gapped superconductors will become arbitrarily small as the temperature is decreased, enabling ultra-high quality factor superconducting resonators for high-efficiency particle accelerators, as initially demonstrated by Schwettman and Turneaure utilizing a solid Nb cavity \cite{Mcas73}.  Materials and infrastructure development over many years has made solid Nb cavities the method of choice for high-efficiency and compact charged particle accelerators,\cite{Pad98} even to the point that normal metal accelerator cavities are now being replaced by their superconducting versions \cite{Gon15}.  \\
	
Experimental basic research efforts have led to many important developments in microwave superconductivity.  The ability to perform measurements at low temperatures were greatly expanded by the availability of the Collins cryostat and abundant quantities of liquid Helium in the late 1940's \cite{Coll47}. In the 21st century world-wide demand for Helium gas and liquid has proven problematic for low-temperature measurements and applications.  A switch to closed-cycle refrigerators has occurred as developments in efficient refrigerator technology have accelerated (see below).  The availability of user-friendly automated dilution refrigerators in the second decade of the current century has helped to greatly expand the experimental and practical use of superconductors at ultra-low temperatures where quantum microwave effects are dominant. \\
	
Many exciting microwave applications have been enabled by the elucidation of the Josephson effects \cite{Jos62} and the experimental demonstration of the AC Josephson effect \cite{Gia65,Yan65} in particular.  It was quickly realized that a Josephson junction acts as a nonlinear and parametric inductor, making it ideal for low-loss parametric amplification of high frequency signals \cite{Feld75}.  The time dynamics of the Josephson junction is ultimately restricted by the plasma frequency of the junction, $\omega_p = 1/\sqrt{L_{JJ} C}$, where $C$ is the capacitance of the junction, so that $f_p=\omega_p/2\pi$ can range from 10's of GHz to 1 THz, depending on the junction size and design.  Hence the junction can act on remarkably short time scales (as small as a few ps), giving rise to quantized voltage pulses, as discussed in the context of RSFQ logic above, among other things.  This, combined with the low-loss and low-dispersion properties of superconducting transmission lines, has led to several generations of Josephson-based digital logic families \cite{Fulton73,Kosh87}. One resulting application is high-speed analog-to-digital (A/D) conversion, which was demonstrated in a sampling oscilloscope that was decades ahead of its time,\cite{Whit87a,Whit87b} while further refinement has achieved A/D sampling rates up to 20 GHz \cite{Muk04}. Related logic families have been used to create entirely superconducting microprocessors in which essentially every function of a computer (including memory) is executed with superconducting circuits \cite{Kot90}.\\
	
Another remarkable high frequency application of the AC Josephson effect is the development of the world-wide voltage standard based on conversion of a microwave frequency into a precisely controlled voltage value, typically either 1 V or 10 V  \cite{Lev77}. This concept has been taken one step further through the creation of a Josephson arbitrary waveform synthesizer \cite{Benz96}.  In this case short current-pulses are sent to an array of Josephson junctions, generating quantized arbitrary waveforms with excellent spectral purity with low noise and no drift \cite{Benz15}.\\
	
The un-diminished superfluid screening response under alternating fields led to the demonstration of superconducting kinetic inductance, an effect long surmised by Helmholtz and Hertz, but not directly observed until the 1960's \cite{Little,Mes69}.  Large kinetic inductance that is extremely sensitive to environmental perturbations have proven a key enabling technology for extremely sensitive detectors of electromagnetic radiation such as MKIDs \cite{Day03,Z12,Oc19}.  At lower frequencies Josephson tunnel junction based detectors provide very high sensitivity to mm-wave and sub-THz radiation \cite{Z04}.  When such radiation is directed onto a junction (typically attached to an antenna), the DC current-voltage curve will exhibit both Shapiro steps as well as steps due to photon-assisted tunneling \cite{Dayem67}. This allows heterodyne detection of radiation with frequencies between roughly 100 GHz and 1 THz.  Both tunnel junctions and MKIDS have the advantage that they can be easily multiplexed into large arrays for imaging.   \\
	
Many superconducting microwave devices based on low-loss, high kinetic inductance, and the Josephson effects were explored in the 20th century using low-transition temperature superconductors \cite{Deaver78}.  The discovery of high-$T_c$ (cuprate) superconductors by Bednorz and M{\"u}ller in 1986 \cite{Bed86} (Nobel Prize in Physics in 1987) had a tremendous impact on microwave superconductivity.  These new materials promised operation above the boiling temperature of liquid nitrogen, and in many cases require operating temperatures that can be easily reached with single-stage closed-cycle cryocoolers.  Many new applications were pursued in the subsequent years, with high-performance microwave bandpass filters being one of the most commercially successful \cite{Simon04,Will01,Will01b}.  The HTS filters make use of carefully coupled high-Q thin film resonators of compact design to produce extraordinarily low-insertion loss and dramatically abrupt band-edge performance \cite{Berk02}.  To illustrate the advanced stage of development of these microwave applications, a spacecraft made up of 8 HTS superconducting microwave devices (including filters, receivers, analog-digital converter, delay line and antenna array) acting as a system was launched into earth orbit \cite{Nis01}.  The system was operated on a space-qualified cryocooler that provided a temperature of 65 K, and the system operated successfully for two years, which was the lifetime of the program.\\

Another remarkable property of HTS cuprate superconductors is their layered structure, with superconductivity being confined mainly to Cu-O layers which are separated from each other by nominally non-superconducting 'spacer materials' in the perpendicular direction.  Associated with this layering is a built-in and naturally occurring Josephson coupling between superconducting layers.  This property was explicitly demonstrated by applying a DC voltage along the perpendicular direction of a strongly-layered single crystal of a cuprate superconductor and observing Josephson radiation from the crystal \cite{Klein92}. This was later engineered into a remarkable voltage-controlled Josephson oscillator that spans the frequency range from 100's of GHz up to over 1 THz \cite{Ozy07}.\\

The ability of superconductors to handle high current densities while maintaining low losses has enabled numerous microwave applications.  Planar (thin film) microwave filters utilizing patterned resonant structures incorporate many compact coupled high-Q resonant structures that create a tailored transmission response with very little insertion loss and steep transmission drop-off out of band \cite{Will01,Will01b,Berk02,Simon04}.  Similar ideas have gone into the development of resonant and non-resonant superconducting metamaterials.\cite{Ricci05,Anl11,Jung14,Laz18}  In this case the meta-atoms are composed of extreme sub-wavelength structures that maintain high-Q despite large microwave fields \cite{Z12}. The incorporation of macroscopic quantum effects into superconducting meta-atoms has made them extraordinarily sensitive to RF and DC magnetic fields, and enabled metamaterials with extreme nonlinearity \cite{Zhang16}.  Utilizing microscopic quantum effects has led to the development of qubit-based superconducting metamaterials that open a new field of research into truly quantum metamaterials \cite{DCL06,Jung14,Macha14}. \\
	
Microwave superconductivity has provided the setting for the revolutionary and rapid rise of superconducting quantum computing and quantum information science (QIS).  Other QIS technologies, such as semiconductor quantum dots, impurity spins, and trapped ions, all depend on RF/microwave signal manipulations and cryogenic technologies.  Landmark results in this rapidly evolving field include the development of the first qubit,\cite{Fried00} circa the turn of the 21st century, the development of a single microwave photon on-demand source,\cite{Houck07} and the development of a sensitive single microwave photon detector \cite{John10,Zhao17}.  QIS measurements are now routinely done in the single microwave photon limit in microwave resonators coupled to microwave transition-frequency qubits \cite{Krantz19}.

\section{MICROWAVE SUPERCONDUCTING INFRASTRUCTURE}
A long-time major limitation for the adoption of superconducting microwave technology has been the issue of cryogenic cooling.  Much of the historical research on superconductivity was performed with liquid cryogens, namely helium and nitrogen, which are consumable materials with inconsistent supply, at least in the case of helium.  However, the advent of closed-cycle mechanical cryocoolers, and numerous derivatives with increasing efficiency and fewer moving parts, has revolutionized cryogenic technology \cite{Ter01,Ter02}.  Today there are many options for low-cost and highly reliable cryocooler technologies, especially those adapted for cryogenic microwave applications \cite{Br06}.  For example, in the last 30 years the pulsed tube refrigerator, which eliminates all moving parts at low temperatures, has greatly increasing reliability without compromising efficiency.\\

Another important issue for microwave superconductivity is the ability to get high frequency signals back and forth to the cryogenic environment without compromising the integrity of the signal or the efficiency of the cryogenic cooling system.  This has led to development of low-microwave loss transmission lines that are simultaneously a small heat-load on the cryogenic environment \cite{Tuk16}.  Examples of such structures include flexible dielectric tapes with an array of superconducting or low-loss metallic transmission line structures.  Superconductors have the advantage of being poor thermal conductors below $T_c$, no worse than insulators in most cases.\\

The QIS revolution has led to a proliferation of commercially-produced and cryogenically-qualified microwave devices \cite{Krin19}.  Examples of such passive devices include attenuators, isolators, circulators, and switches.  Active devices have recently seen great advancement in terms of low-dissipation broadband low-noise amplifiers that are very well suited for the cryogenic environment.  There has also been great leaps forward in development of quantum-limited amplifiers (QLA) based on the parametric properties of Josephson junctions.  Two main classes of QLAs are currently in use, Josephson traveling wave parametric amplifiers and Josephson parametric converters \cite{Aum20}.\\

Finally, as noted above, there has been development of new cryogenic microwave sources and detectors capable of operating down to the single photon limit.  Both the intrinsic Josephson effect in cuprates, and the use of artificial Josephson devices, have enabled these new technologies. \\

\section{THE FUTURE}
	It is clear that the quantum information revolution is built squarely on the foundation of microwave superconductivity.  Tremendous microwave engineering challenges are in store for the development of large scale quantum coherent computing machines, creating many opportunities for new applications of superconductors.   \\
	
	It seems likely that new superconductors with interesting properties will continue to be discovered.  Most technologically-relevant superconductors are s-wave superconductors, meaning that the two electrons that make up a Cooper pair enter into an $\ell = 0$ quantum angular momentum state.  These superconductors generally have a full excitation energy gap on the Fermi surface, giving rise to an exponentially small number of quasiparticles in the limit of zero temperature, among other features.  An increasing number of superconductors discovered since the 1970's have shown clear evidence of $\ell = 1$ and $\ell = 2$, and possibly higher, quantum angular momentum pairing states of the electrons.  These materials generally have \textit{nodes} in their energy gap, meaning that quasiparticles can be excited even at the lowest temperatures, thus creating altogether different low-energy properties of these materials.  So far, few of these more exotic materials have found an application based specifically on these pairing properties.  Related to this, superconductors with non-trivial electronic topological properties have recently been proposed and discovered \cite{Deng16,Kallin16}.  These materials may host chiral edge currents and perhaps other exotic phenomena that may find use in future QIS applications \cite{Sato17,Jiao20}. \\ 
	
	It is also likely that new superconductors with transition temperatures exceeding 100 K will continue to be discovered.  So far many of these higher-$T_c$ materials have been difficult to utilize in applications because of their toxic chemical constituents, brittle mechanical properties, or the fact that they can only be stabilized under extraordinarily high pressures.  However, our ability to predict the properties of new materials, and their stability, is growing more sophisticated with time \cite{Liu17,Peng17}.  This theoretical effort has directly led to discovery of new superconductors with transition temperatures approaching room temperature,\cite{Droz19,Som19} the holy grail of the superconducting materials community \cite{Beas11}.  Turning these new materials into practical devices and products will take time, but it seems likely that superconducting microwave devices will find increasingly wider application and usage in the future.\\
	
\section{CONCLUSION}
	This broad overview of microwave superconductivity is intended to give the reader a taste of this very exciting field of microwave technology that shows great promise for young engineers and technologists.  Harnessing the unique quantum mechanical properties of the superconducting state offers many opportunities for invention and for the solution to numerous problems in modern life.  We hope that this review will help to inspire more creative uses of superconductors in microwave devices, systems, and applications.
	
\section*{ACKNOWLEDGMENT}
I thank the many advisors, colleagues, current and former students and post-docs who have contributed to my understanding of microwave superconductivity.  Specifically I would like to thank Bakhrom Oripov for performing the TDGL Meissner calculation and creating Fig. \ref{MeissnerFig}, and Alexander P. Zhuravel for providing Fig. \ref{LSMPR}.  I also thank Stuart Berkowitz, Marty Nisenoff and Dan Oates for helpful comments.  This work was supported by the US Department of Energy through grant DESC0018788 (support of SMA) and the US National Science Foundation through grant number DMR2004386.


\bibliographystyle{ieeetr}
\bibliography{Anlage_IEEE_Microwaves_Microwave_Superconductivity_Bibliography}

\end{document}